\documentclass{PoS}

\usepackage{url}

\title{VHE Observations of Galactic binary systems with VERITAS}

\ShortTitle{VHE Observations of Galactic binary systems with VERITAS}

\author{\speaker{Gernot Maier} for the VERITAS Collaboration\thanks{full author list at https://veritas.sao.arizona.edu/}\\
        DESY, Platanenallee 6, D-15738 Zeuthen, Germany\\
        E-mail: \email{gernot.maier@desy.de}}

\abstract{
Gamma-ray binaries are variable sources of GeV-to-TeV photons with the peak in their spectral energy distributions above 100 MeV.
There are only five Galactic gamma-ray binaries known and the processes which govern particle acceleration, non-thermal emission and the variability patterns are not well understood.
We present here for the first time results of the VERITAS binary discovery program at VHE energies and a summary of eleven years of observations of the gamma-ray binary HESS J0632+057. 
VERITAS has acquired more than 220 hours of observations on HESS J0632+057. The results are discussed in the context of contemporaneous  observations with Swift XRT at X-ray energies.}

\FullConference{35th International Cosmic Ray Conference\\
		 12-20 July, 2017\\
		 Bexco, Busan, Korea}

\begin{document}

\section{Introduction}

The small class of gamma-ray emitting galactic binary system consists currently of five systems: PSR B1259-63, LS 5039, LS I +61 303, HESS J0632+057 and 1FGL J1018.6-5856 (see \cite{Dubus:2013a} for a review).
Additionally, the TeV source TeV J2032+4130, associated with the gamma-ray pulsar PSR J2032+4127, was recently reported to be possibly  part of a very eccentric, long orbital period binary system  \cite{Ho-2017}.
All gamma-ray binaries show variable, very-high energy gamma-ray emission ($>$ 100 GeV), for which the variability in most cases can be associated with the orbital motion of the two constituents of the binary systems around their common center of mass.
The systems are composed of a compact object (neutron star or black hole) and a massive star of O or B(e) type.
The spectral energy distributions (SEDs) of gamma-ray binaries peak at energies above 100 MeV, in contrast to the much larger number of known X-ray binaries observed with peaks in the SEDs at significantly lower energies.

The dominant emission mechanism for the non-thermal emission and the physical processes powering the underlying population of high-energy particles are not entirely understood. 
Current observational results do not allow us to distinguish between the two major scenarios for gamma-ray emission, into which the large number of theoretical models published can be divided. In the microquasar model, charged particles are accelerated in an accretion-driven relativistic jet, similar to the processes observed in active galactic nuclei. 
Alternatively, particles are accelerated in the strong shock interaction between the stellar wind of the massive star and an energetic plasma wind powered by the rotation of a pulsar.  
The compact object is known for only one binary system, the pulsar-binary PSR 1259-63.

Flux modulation across the electromagnetic spectrum is observed in all gamma-ray binaries on the time scale of the orbital periods (typically between a few days and several years), but also on much shorter and longer time scales.
As an example, the emission pattern of the binary LS I +61 303 is strongly modulated by the orbital period of 26.5 days (e.g.~\cite{Albert:2009, Acciari:2011} ), but also shows day-long flares \cite{Archambault:2016} and possibly a super-orbital variability pattern on a scale of 1667 days \cite{Ahnen:2016}.
The binary HESS J0632+057 is another example, with a variability pattern consisting of two distinct maxima and 
a short phase of quietness compared to the orbital period \cite{Aliu:2014}.
These variability patterns are considered to be one of the keys for a better understanding of of the emission processes. 
Variability depends on the orbital movement, the geometry of the orbit, the state and extension of the equatorial outflow observed in Be stars, the clumpiness of the wind of the massive star, and for the microquasar model on the variability related to the state of the accretion-fed jet.

The total number of gamma-ray binaries in our Galaxy is not very well known.
The lower bound on the population is given by the number of detected objects with ground- and space-based gamma-ray observatories.
The upper bound on the number of objects is not well constrained by observations, mainly due to the non-detection of HESS J0632+057 in the sub-GeV range by the {\em Fermi}-LAT, despite its  almost continuous monitoring of the sky since 2008 \cite{LAT:2017}.
The number of gamma-ray binaries allows constrains to be put on the population of bound binaries after supernova explosions, as young pulsars are required to produce the strong shocks required  for efficient particle acceleration in the interaction with the wind of the massive stars (assuming the pulsar-stellar wind acceleration model).

The binary program of the VERITAS gamma-ray observatory is motivated by the limitations on these measurements.
The  program consists of deep observations of known binaries observable in the Northern Hemisphere (LS I +61 303 and HESS J0632+057) and of a discovery program with the aim to increase the number of known gamma-ray binaries.
We present in these proceedings a summary of the VERITAS binary discovery program and  results of eleven observations seasons with VERITAS  of the binary HESS J0632+057.
The results of equally long observations of the binary LS I +61 303 are presented elsewhere at this conference \cite{Kar:2017}.

\section{The VERITAS observatory}

The Very Energetic Radiation Imaging Telescope Array System VERITAS
 is a ground-based gamma-ray observatory built to observe high-energy photons
in the energy range from 85 GeV to $>$30 TeV. 
The VERITAS observatory is located at the Fred Lawrence Whipple observatory in southern Arizona, USA
(1.3 km above sea level, N31$^{\mathrm{o}}$40', W 110$^{\mathrm{o}}$57').
It consists of four 12-meter diameter imaging atmospheric Cherenkov telescopes which observe
simultaneously the faint Cherenkov light that is emitted when particle showers initiated by the impact
of high-energy gamma rays pass through the atmosphere.
The flux sensitivity of the instrument reaches 1\% of the flux of the Crab Nebula in 25 hours
 (assuming a detection significance of 5 standard deviations).
All observations presented here are taken under dark conditions.
For more information on VERITAS, see \url{http://veritas.sao.arizona.edu/}.

\section{The VERITAS Binary Discovery Program}

The discovery of a new gamma-ray binary could provide significant progress in the understanding of the underlying physical processes describing acceleration and high-energy emission in these systems.
The VERITAS discovery program aims to perform a systematic search for new gamma-ray binaries in our Galaxy.
Fourteen objects have been selected from different lists of high-mass X-ray binaries \cite{Liu:2007, Reynolds:2013, Raguzova:2007} according to the following criteria: a massive companion star of O or B(e)-type, a relatively small distance ($<$5 kpc), and a preference for orbital periods $<50$ days for those objects with known orbital parameters.
It should be noted that none of the observed binaries has been suggested to be a high-energy gamma-ray emitter; this is not surprising, given the little existing understanding of the acceleration and emission process in these systems.

Observations with VERITAS were spread out over the entire orbital period for these objects, resulting in sequences of short (30-90 min) exposures for each night spread over several months. 
A total of typically 10-15 observing hours per object were taken, corresponding to a sensitivity in units of the flux of the Crab Nebula above 300 GeV of 1.5-2\%.
The binary discovery program is part of the VERITAS long-term observing plan; we report here first results of observations on four binary systems:

{\bf IGR J00370+6122:} This bright Integral source consists of a B-star and a compact object (likely a neutron star) in an orbit of 15.667 days \cite{Grunhut:2014}. 

{\bf XTE J0421+560:} The X-ray transient XTE J0421+560 was discovered during a bright outburst by the All-Sky Monitor on-board RXTE.
The massive star is of Be type; the proposed orbital solution (period of $19.41 \pm 0.02$ days)  puts the periastron distance close to the surface of the massive star \cite{Bartlett:2013}.    

{\bf V662 Cas:} The wind-accretion, supergiant, high-mass X-ray binary consists of a massive B1Ia star and a neutron star.
Three  periodicities are detected from V662 Cas: a 9700 s neutron star rotation period, an 11.6 day orbital period, and a 30.7 day superorbital modulation \cite{Corbet:2013}.
The system is located at a distance of $5.9\pm1.4$ kpc

{\bf LS V +44 17:} This Be X-ray binary pulsar system is located at a distance of $3.3\pm0.5$ kpc and consists of a B0.2Ve star and a pulsar with a pulsation period detected in X-rays of $202.5\pm 0.5$ s \cite{Reig:2005}.
The orbital parameters are estimated to be $150\pm0.2$ days, estimated from Swift XRT observations \cite{Ferrigno:2013}.

\begin{table}[ht]
\centering
\begin{tabular}{l|c|c|c|c}

{Binary} &  {Observing } & {Elevation } & Significance & {Flux upper limits} \\
&   Time  &Range & [$\sigma$] & (99\% confidence level) \\
&         [minutes]        & & & [cm$^{-2}$ s$^{-1}$] \\
\hline
IGR J00370+6122 & 1009 & $52^{\mathrm{o}} - 61^{\mathrm{o}}$ & 2.8 & $3.6 \times 10^{-13}$ \\
XTE J0421+560 & 1466 & $50^{\mathrm{o}} - 66^{\mathrm{o}}$ & -0.7 & $5.2\times 10^{-13}$ \\
V662 Cas &  694 & $52^{\mathrm{o}} - 56^{\mathrm{o}}$ & 0.1 & $1.2\times 10^{-12}$ \\
LS V +44 17 &  849 & $48^{\mathrm{o}} - 77^{\mathrm{o}}$ & -2.7 & $3.1\times 10^{-13}$ \\
\end{tabular}
\caption{Summary of VERITAS observations of four high-mass X-ray binary systems.
Flux upper limits are given for energies $>350$ GeV.}
\label{tab:binaries}
\end{table}

Table \ref{tab:binaries} summarises the results of the VERITAS observations.
None of the binaries has been detected at energies above 350 GeV.
Typical upper flux limits of 1-2\% of the flux of the Crab Nebula have been derived.

\section{Long-term monitoring of the gamma-ray binary HESS J0632+057}

The gamma-ray binary HESS J0632+057 (VER J0632+057) was discovered serendipitously by the High Energy Stereoscopic System (H.E.S.S.) during observations of the Monoceros Loop supernova remnant \cite{Aharonian:2007}.
VERITAS observations revealed that HESS J0632+056 is a point-like and variable source of very-high energy ($>$ 100 GeV) gamma rays \cite{Acciaro:2009}.
The system consists of a massive Be star (MWC 148/HD 259440) and a compact object of unknown type (neutron star or black hole)
 located at a distance of 1.1-1.7 kpc \cite{Aragona:2010}.
Long-term X-ray observations using the Swift X-ray Telescope (XRT) revealed periodical flux modulation, pointing towards the binary nature of the system \cite{Bongiorno:2011}.
The long orbital period of the binary of $315\pm5$ days \cite{Aliu:2014} makes it difficult to perform deep and continuous observations of the system during all orbital phases.
Further gamma-ray observations by VERITAS, H.E.S.S. \cite{Aliu:2014} and MAGIC \cite{Aleksic:2012} revealed a pattern of variability, with significant gamma-ray emission above 1 TeV in two different phase ranges of the orbit.  
HESS J0632+057 is the only gamma-ray binary which has not been detected at MeV-GeV energies with the Fermi LAT \cite{Caliandro:2013}.
We present here new measurements of the gamma-ray binary with the VERITAS telescopes obtained during observations in season 2016-2017 and summarise the eleven years of VERITAS observations on this system.

\begin{figure}
 \centering
     \includegraphics[width=.8\textwidth]{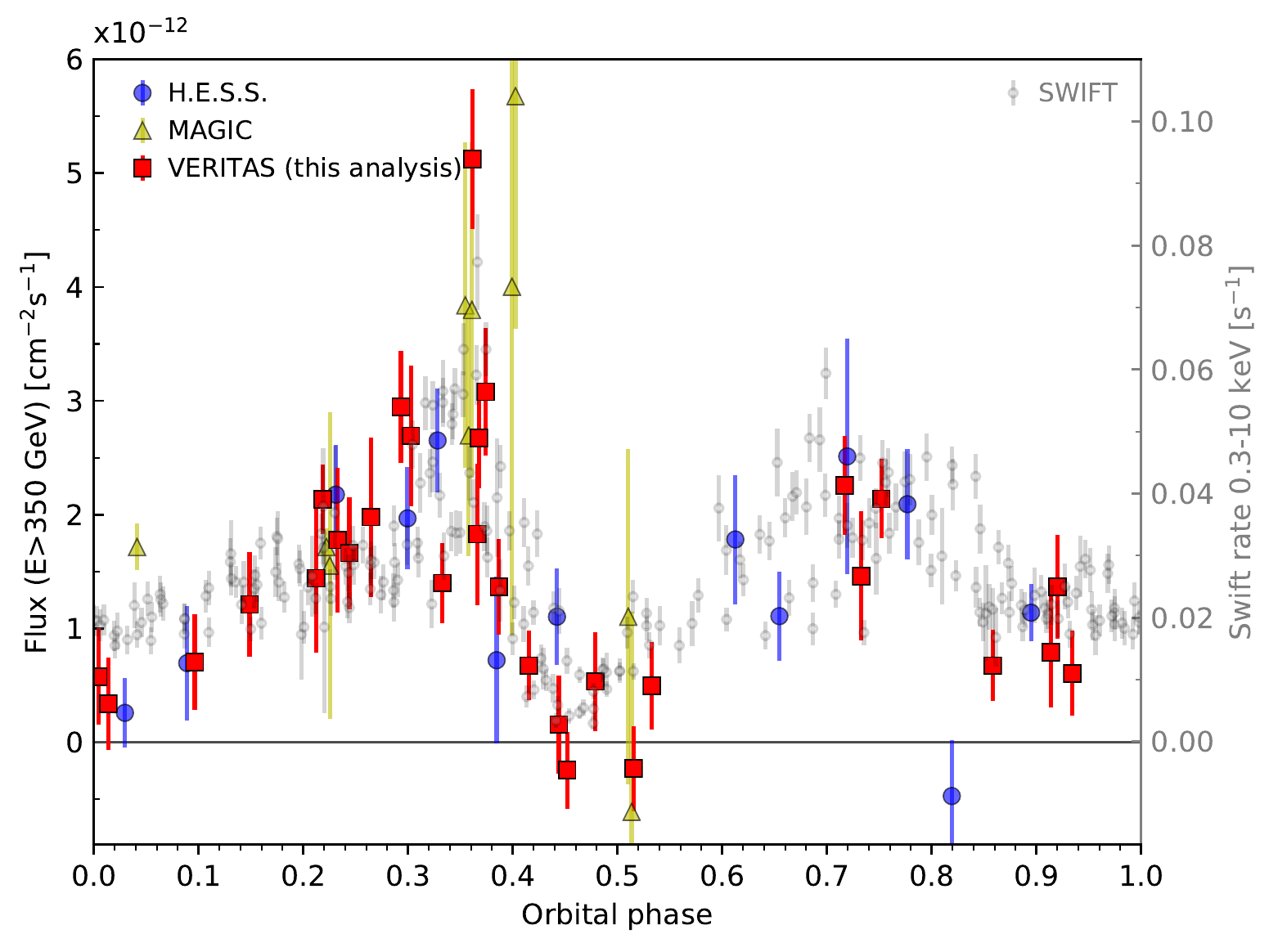}
     \caption{Phase-folded Long-term gamma-ray observations of HESS J0632+057 with VERITAS (this paper), H.E.S.S. \cite{Aliu:2014}, and MAGIC \cite{Aleksic:2012} at energies >350 GeV and X-ray observations with Swift-XRT (0.3-10 keV; gray markers).
 An orbital period of 315 days was assumed for the phase folding.}
     \label{fig1}
  \end{figure}

HESS J0632+057 has been observed by VERITAS for a total of 220.5 hours between 2006 December and 2017 March at energies above 350 GeV, see Figure \ref{fig1}. 
The binary has been detected  by VERITAS with a total significance of 21.3$\sigma$.
At X-ray energies, Swift XRT monitored HESS J0632+057 at 0.3-10 keV from 2009 January to 2017 March

\begin{figure}
 \centering
     \includegraphics[width=.45\textwidth]{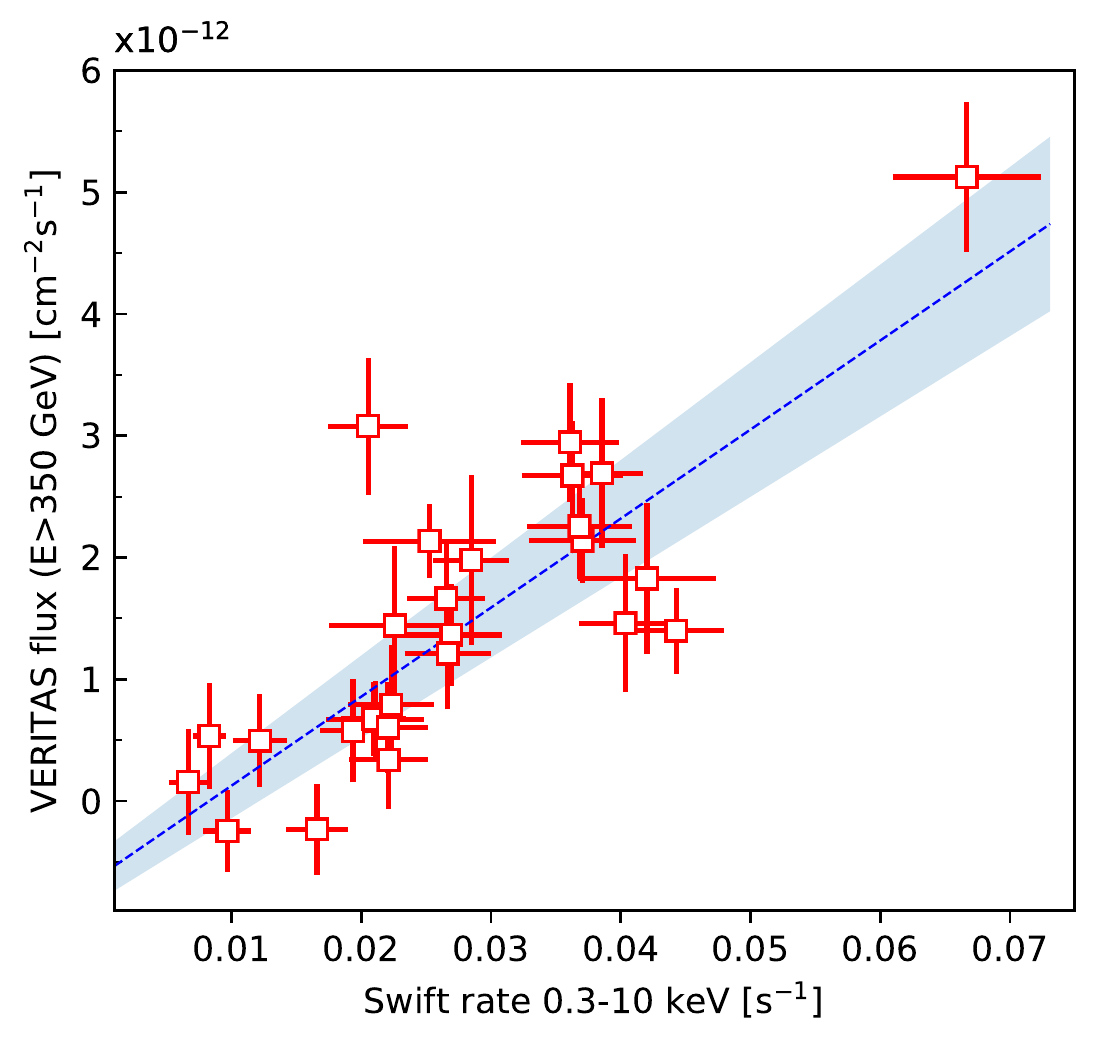}
      \includegraphics[width=.45\textwidth]{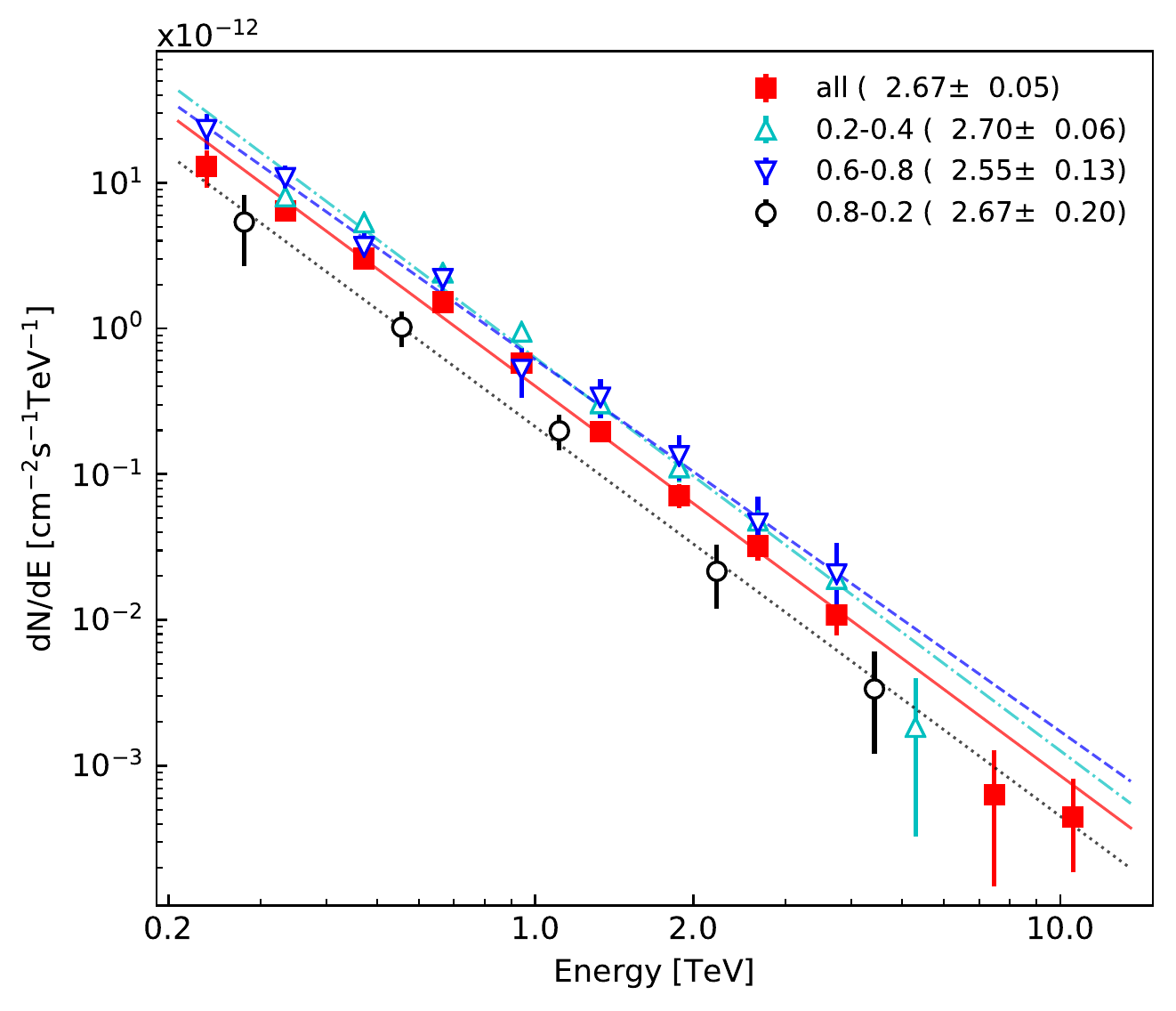}
     \caption{Left: Gamma-ray ($>350$ GeV) fluxes vs. X-ray (0.3-10 keV) rates for contemporaneous observations (defined here as $\pm 2.5$ day intervals around the date of the gamma-ray observations). 
     The blue line indicates a linear fit to illustrate the correlation of the fluxes at the two different energies.
      Right: Differential energy spectra for gamma-ray photons averaged over all orbital phases (red markers and red line), and for the orbital phase ranges 0.2-0.4 (green markers and green line), 0.6-0.8 (blue markers and blue line), and 0.8-0.2 (black markers and black line). The lines show results from fits assuming power-law distributions. Fit results can be found in Table \ref{table:results}.}
     \label{fig2}
  \end{figure}

Figure \ref{fig1} shows that the gamma-ray emission from HESS J0632+057 is clearly variable with two distinct maxima.
The first maximum at phases 0.2-0.4 is brighter than the emission at phases 0.6-0.8, but like the X-ray measurement, exhibits flux variability at similar phases. 
The binary is clearly detected in phases 0.2-0.4, 0.6-0.8, and 0.8-0.2 (see Table \ref{table:results}).
It was suggested \cite{Casares:2012}  that the orbit is eccentric ($e\approx0.83$) using radial velocity measurements of MWC 148. 
This puts the largest maximum of the gamma-ray emission at about 0.3 phases after the periastron, similar to the binary LS I +61 303 (see e.g.~\cite{Acciari:2008}).
The variability pattern of the phase-folded gamma-ray light curve follows the X-ray light curve.
The correlation between the flux at the two energies ranges is 0.816 (Pearson correlation, p-value $3.78\times 10^{-7}$), see Fig \ref{fig2}, left.

\begin{table}[ht]
\centering
\begin{tabular}{l|c|c|c|c|c}
{ Orbital phase} & { all phases} & {0.2-0.4}  & {0.4-0.6} & { 0.6-0.8} &{  0.8-0.2} \\
\hline
{Observation time} (h) & 220.5 & 88.5 & 45.1 & 26.9 & 60  \\
{ Significance} ($\sigma$) & 23.1 & 19.7 & 2.7 & 11.4 & 6.2 \\
\hline 
{Flux Normalization} $\Phi_0$  &  & & & & \\
{\ at 1 TeV} & $4.1\pm0.2$ & $6.4\pm0.4$ & - & $6.2\pm0.7$ & $2.1\pm0.4$ \\
{ Photon index} $\gamma$ & $2.67\pm0.05$ & $2.70\pm0.06$  & - & $2.55\pm0.13$  & $2.67\pm0.2$ \\
$\chi^2$/N & 15.2/8 & 22.8/7 & - & 2.7/7 & 0.8/3 \\
\end{tabular}
\caption{
\label{table:results}
Results of the spectral analysis results for energies $>$350 GeV for the phase-folded VERITAS measurements.
A power-law distribution $dN/dE = \Phi_0 \cdot E^{-\gamma}$ of the data is assumed for the spectral fits,
see also Figure \ref{fig2} (right).
All quoted errors are $1\sigma$ statistical errors only.
The flux normalisation constant  $\Phi_0$ is in units of $10^{-13}$ cm$^{-2}$ s$^{-1}$ TeV$^{-1}$.
Systematic uncertainties on the flux normalisation are typically of the order of 20\%.
}
\end{table}

The differential energy spectra in gamma rays during the two maxima and for the phase range 0.8-0.2 can be described by power-law distributions (Fig \ref{fig2} (right) and Table \ref{table:results}).
The slope parameters of the spectral fits are consistent with each other, which points towards similar physical conditions at the gamma-ray emission sites during the high states before and after the apastron phase.

\section{Conclusions}

The VERITAS long-term observation plan includes both deep observations of the known gamma-ray binaries LS I +61 303 and HESS J0632+057 and a discovery component consisting of observations of binaries with similar properties as the known ones.
We have presented in these proceedings new observations of four gamma-ray binary candidates, which have resulted in no new detections.
The VERITAS results on HESS J0632+057 consist of flux and precise spectral measurements of the gamma-ray emission at all orbital phases.
The binary stands out among the five known gamma-ray binaries with its two maxima in the phase-folded light curve, and a sharp minimum in the emission pattern close to the apastron phase.
Gamma- and X-ray fluxes are highly correlated. 
A better measurement of the geometrical parameters describing the system and the properties of the stellar disk (extension, orientation, density) are very likely necessary for a better understanding of the light curve pattern in HESS J0632+-57.
The VERITAS observations of binaries will continue for the coming years.

\acknowledgments

This research is supported by grants from the U.S. Department of Energy Office of Science, the U.S. National Science Foundation and the Smithsonian Institution, and by NSERC in Canada. We acknowledge the excellent work of the technical support staff at the Fred Lawrence Whipple Observatory and at the collaborating institutions in the construction and operation of the instrument. The VERITAS Collaboration is grateful to Trevor Weekes for his seminal contributions and leadership in the field of VHE gamma-ray astrophysics, which made this study possible. GM acknowledges the support by the Helmholtz Association.

This work made use of data supplied by the UK Swift Science Data Centre at the University of Leicester.


\begin{thebibliography}{99}
\bibitem{Dubus:2013a} Dubus, G. 2013, Astronomy and Astrophysics Review, 21, 64
\bibitem{Ho-2017} Ho, W. C. G. et al. \ 2017, MNRAS, 464, 1211
\bibitem{Albert:2009} Albert, J. et al. \ (MAGIC Collaboration), 2009, ApJ, 693, 303
\bibitem{Acciari:2011} Acciari, V. et al. \ (VERITAS Collaboration), 2011, ApJ, 738, 3
\bibitem{Archambault:2016} Archambault, S. et al. \ (VERITAS Collaboration), 2016, ApJ, 817, L7
\bibitem{Ahnen:2016} Ahnen, M.L. et al. \ (MAGIC Collaboration), 2016, A\&A, 591, A76
\bibitem{Aliu:2014} Aliu, E. et al.\ (VERITAS \& H.E.S.S. Collaborations), 2014, ApJ, 780, 168
\bibitem{LAT:2017} Fermi-LAT Collaboration, arXiv:1702.00664
\bibitem{Kar:2017} Kar, P. et al. \ (VERITAS Collaboration), 2017, these proceedings
\bibitem{Liu:2007} Liu, Q.Z. et al. 2007, A\&A, 469, 807
\bibitem{Reynolds:2013} Reynolds, M.T. \& Miller, J.M. 2007, ApJ, 769, 16
\bibitem{Raguzova:2007} 	Raguzova, N. V. \& Popov, S. B. 2005, A\&AT, 24, 151
\bibitem{Grunhut:2014} Grunhut, J.H. et al. \ 2014, A\&A, 563, A1 
\bibitem{Bartlett:2013} Bartlett, E.S. et al. 2013, MNRAS, 429, 1213
\bibitem{Corbet:2013} Corbet, R. \& Krimm, H. 2013, ApJ, 778, 45
\bibitem{Reig:2005} Reig, P., Negueruela, I., et al. 2005, A\&A, 440, 1079
\bibitem{Ferrigno:2013} Ferrigno, C., Farinelli, R., et al. 2013, A\&A, 553, 103
\bibitem{Aharonian:2007} Aharonian, F. et al. (H.E.S.S. collaboration), 2007, A\&A, 469, L1
\bibitem{Acciaro:2009} Acciari, V.A. et al. (VERITAS collaboration), 2009, ApJ 698, L94
\bibitem{Aragona:2010} Aragona, C., McSwain, M. V., \& De Becker, M. 2010, ApJ, 724, 306
\bibitem{Bongiorno:2011} Bongiorno, S.D. et al., ApJ 2011, 737, L11
\bibitem{Aleksic:2012} Aleksi\'{c}, J. et al.\  (MAGIC Collaboration) 2012, ApJ, 754, L10
\bibitem{Caliandro:2013} Caliandro, G.A. et al. \ 2013, MNRAS 436, 740
\bibitem{Casares:2012} Casares, J. et al. \ 2012, MNRAS 421, 1103
\bibitem{Acciari:2008} Acciari, V.A. et al (VERITAS collaboration) 2008, ApJ 679, 1427



\end{thebibliography}
\end{document}